\documentstyle[epsf,prb,floats,twocolumn,aps]{revtex}

\draft
\begin{document}

\title{Unusual features in the nonlinear microwave surface impedance
of YBaCuO thin films}
\author{A.P.Kharel$^1$, A.V.Velichko$^{1,2}$, J.R.Powell$^1$, A.Porch$^1$,
M.J.Lancaster$^1$ and R.G.Humphreys$^3$}
\address{$^1$School of Electronic and Electrical Engineering, The
University of Birmingham, B15 2TT, UK}
\address{$^2$Institute for Radiophysics and  Electronics of NAS,
Kharkov, Ukraine}
\address{$^3$DERA, St. Andrews Road, Malvern WR14 3PS, UK}

\date{\today}

\maketitle

\begin{abstract}
Striking features have been found in the nonlinear microwave (8.0 GHz)
surface impedance $Z_s=R_s+j\cdot X_s$ of high-quality
$Y\!BaCuO$ thin films with comparable low power characteristics
($R_{res}\sim $35--60~$\mu\Omega$ and $\lambda_L(15~K)\sim $130--260~nm).
The surface resistance $R_s$ is found to increase, decrease or remain
independent of the microwave field $H_{r\!f}$ (up to 60~mT) at different
temperatures and for different samples. However, the surface reactance
$X_s$ always follows the same functional form. Mechanisms which may be
responsible for the observed variations in $R_s$ and $X_s$ are briefly
discussed.
\vspace*{0.3 true cm}%
\end{abstract}


Measurements of the nonlinear microwave surface impedance, $Z_s$, of
high-temperature superconductors (HTS) is a powerful tool for studying
non-equilibrium processes in these materials.  Nonlinear impedance
measurements allow one to investigate peculiarities of the $r\!f$-vortex
nucleation, and to study the vortex dynamics at elevated microwave fields.
Such measurements may also discriminate between d-wave and
s-wave mechanisms of pairing symmetry in HTS, and
indicate the presence of magnetic impurities in the
materials~\cite{Kres1,Hein}.

   In the present paper, we report observations of non-monotonous
behavior of $R_s$ and the penetration depth, $\lambda$  (or, equivalently,
the surface reactance $X_s=\omega\mu_0\lambda$), of high-quality epitaxial
$Y\!BaCuO$ thin films, in microwave fields up to 60~kA/m ($\sim $700~Oe)
using the coplanar resonator technique~\cite{Porch1} at 8~GHz.
For all samples, depending on temperature $T$, $R_s$
demonstrates completely different behavior, whereas
$\lambda$ always preserves the same $H_{r\!f}$-dependence, irrespective of
sample and $T$.  Measurements are presented for very high quality samples
over a wide temperature range (12--75~K) which at first time reveal {\it
non-monotonous\/} and {\it uncorrelated behavior\/} in $R_s$ and $X_s$ as
a function of $H_{r\!f}$.  Such a behavior does not agree with any of the
existing models for the nonlinear microwave
impedance~\cite{Oates2,Golos2,Sridh,Herd2,Halbr5,Belk}.  In the
following we discuss several mechanisms relevant to these observations.

The films are deposited by e-beam co-evaporation onto polished
(001)-orientated  MgO single crystal substrates $10\times10$~mm$^2$. The
films are 350 nm thick. The c-axis misalignment of the films are typically
less than 1$\%$,  and the $dc$ critical current density $J_c$ at 77~K is
around $2\cdot10^6$ A/cm$^2$.  More detailed information on the growth
technique can be found in~Ref.[\onlinecite{Chew}].
The values of $R_s$ and $\lambda$ at 15~K are 60, 35, 50~$\mu\Omega$  and
260, 210, 135~nm for samples TF1, TF2 and TF3, respectively.

\begin{figure}[t]
\def\epsfsize#1{0.35}
\vspace*{-2.0 true cm}
\centerline{\epsfbox{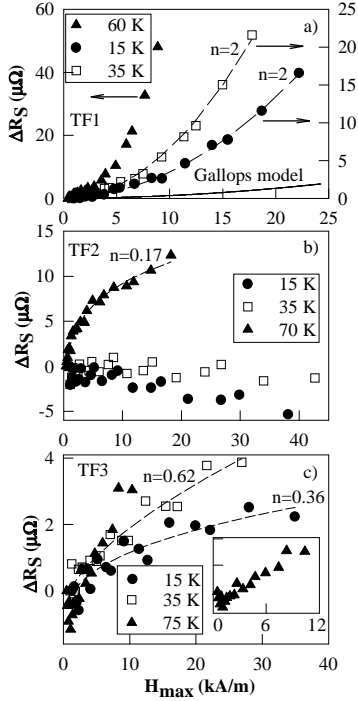}}
\vspace*{-2.0 true cm}
\caption{Microwave field $H_{r\!f}$ dependences of the change in the
surface resistance $\Delta R_s$ for three samples TF1, TF2 and TF3 at
different temperatures $T$. The dashed lines are fitting curves using a
function $\sim H_{r\!f}^n$, which is discussed in the text. $T$- and
$n$-values are given in the figure. The solid line in a) is a fit using
the modified model of Gallop et al.\ ref[14]. The parameters of the
fit are as follows: normal resistance $R_n=1.83$~$\Omega$, zero field
critical current density $J_{c0}=10^{10}$~A/cm$^2$, grain size
$\alpha=6.4\cdot 10^{-7}$~m, grain penetration depth
$\lambda_{ab0}=2.51\cdot 10^{-7}$~m. The insert shows the data at 75~K
on an expanded scale.}
\label{fig1}
\end{figure}

\begin{figure}[t]
\def\epsfsize#1{0.35}
\vspace*{-2.0 true cm}
\centerline{\epsfbox{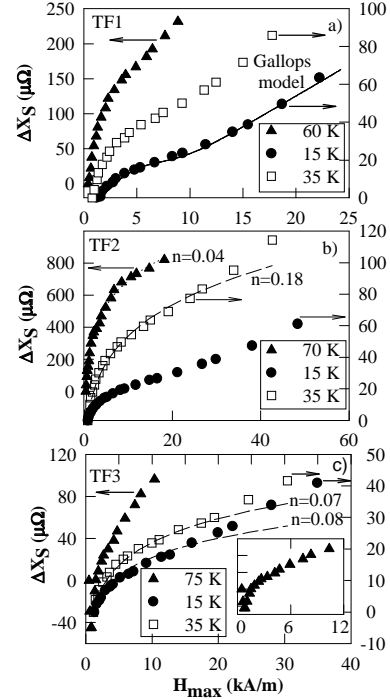}}
\vspace*{-2.0 true cm}
\caption{The change in the surface reactance $\Delta X_s$ as a function of
$H_{r\!f}$ for three samples TF1, TF2 and TF3 at different temperatures.
The fits are the same as those plotted in fig.1. The insert shows the data
at 75~K on an expanded scale.}
\label{fig2}
\end{figure}

Changes in $R_s$ and $X_s$ with $H_{r\!f}$, $\Delta
R_s=R_s(H_{r\!f})-R_s(0)$ and $\Delta X_s=X_s(H_{r\!f})-X_s(0)$, are
plotted in fig.~\ref{fig1} and fig.~\ref{fig2} for all three samples.
For sample TF1 for all $T$ and in
the whole field range  $\Delta R_s\sim H_{r\!f}^2$, whereas
for samples TF2 and TF3 the behavior of $\Delta R_s(H_{r\!f}$) changes
dramatically with $T$. For sample TF2 $R_s$ changes from
decreasing at 15 K to almost $H_{r\!f}$-independent behavior at 35 K, and
to a rapidly increasing function of $H_{r\!f}$ between 40--75 K.
At 15 K $\Delta R_s$ diminishes noticeably only at $H_{r\!f}>10$~kA/m
showing no features of saturation up to the highest available $H_{r\!f}$
of $\sim$40~kA/m. At higher $T$ (70~K) a transition to a characteristic
sublinear field dependence ($\sim H_{r\!f}^n$, $n<1$) occurs.
A similar behavior is also observed for sample TF3 at 15 and 35~K over the
whole field range, whereas at $T>70$~K (see fig.~\ref{fig1}c)
a minimum in  $\Delta R_s(H_{r\!f})$ at low fields appears, after which
the usual sublinear $H_{r\!f}$-dependence is recovered.
As regards to $\Delta X_s(H_{r\!f})$, it is always a sublinear function of
$H_{r\!f}$ at low fields with a characteristic kink and superlinear
$H_{r\!f}$-dependence ($\sim H_{r\!f}^n, n>1$) at higher fields (see
fig.~\ref{fig2}).  This dependence of $\Delta X_s$ on $H_{r\!f}$ persists
for all samples and for almost all temperatures, and in general {\it no
correlation is observed\/} between $\Delta X_s(H_{r\!f})$ and $\Delta
R_s(H_{r\!f})$. The only exception is sample TF3 for which
$\Delta X_s(H_{r\!f})$ qualitatively correlates with $\Delta
R_s(H_{r\!f})$ at all $T$, and even the minimum at low fields
is reproduced in both dependences at 75~K (see fig.~\ref{fig1}c and
fig.~\ref{fig2}c). In fig.~\ref{fig1} and fig.~\ref{fig2} some of $\Delta
R_s$ and $\Delta X_s$ data are fitted to the function $\sim H_{r\!f}^n$
which is predicted by Halbritter's model of Josephson vortex motion in
weak links (WL)~\cite{Halbr5} ($0.5<n<2$) and the Ginzburg-Landau theory
for the pair breaking mechanism ($n=2$).

\begin{figure}[t]
\def\epsfsize#1{0.35}
\vspace*{-2.0 true cm}
\centerline{\epsfbox{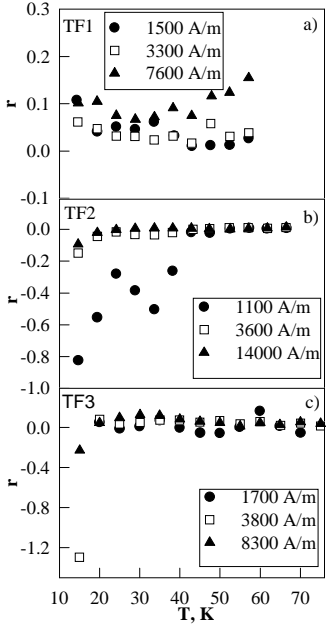}}
\vspace*{-2.0 true cm}
\caption{Temperature dependences of the $r$-parameter ($\Delta R_s/\Delta
X_s$) for samples TF1, TF2 and TF3 at different $H_{r\!f}$ (specified in
the figure).}
\label{fig3}
\end{figure}

In fig.~\ref{fig3}, we plot the temperature dependence of the
$r$-parameter ($r=\Delta R_s/\Delta X_s$) for all the samples at three
microwave power levels. These data are often used to distinguish
between various nonlinear mechanisms~\cite{Golos2,Halbr5}. The general
trend of $r(T)$ for all the samples is a decrease in the absolute value of
$r$ with increasing $T$, gradually saturating at high $T$. For samples
TF2 and TF3 the most pronounced change in $r$ occurs at low $T$, where it
has a large negative value of about $-1$ and rapidly levels off with
temperature approaching a value of 0.01--0.06 (see fig.~\ref{fig3}b,c).
One can see that the initial negative value of $r$ is reduced with
enhanced power. Unlike other samples, TF1 in the high field regime
($H_{r\!f}\sim$ 7600~A/m) displays an increase in the $r$-parameter at
high temperatures ($T>45$~K) and reaches a value of $\sim $0.15. In
addition, the initial low-$T$ $r$-value for sample TF1 depends
non-monotonously on the microwave field (see fig.~\ref{fig3}a).

For explanation of the observed non-monotonous field dependences of
$\Delta R_s$ and $\Delta X_s$ we involve three
different mechanisms.  Each mechanism is capable of describing only
particular features in $H_{r\!f}$-dependences of $\Delta R_s$ and
$\Delta X_s$.
First, the modified weakly-coupled grain
model~\cite{Hylton1}, proposed by Gallop et al.~\cite{Gallop1}, assumes
that for high-quality HTS films a WL between two superconducting grains is
shunted by another third grain, which serves as an additional path for
both $dc$ and $r\!f$ currents. This model presents a highly simplified
picture (not least, because it makes no distinction between the Meissner
and the mixed states). Nevertheless it enables one to reproduce a
reduction of $R_s$, such as we observe, given a certain set of the
material parameters.  However, it is unable to describe two important
features observed by us; sublinear $\Delta X_s(H_{r\!f})$ dependence at
low fields with a characteristic change in curvature at higher $H_{r\!f}$,
and the decrease of $\Delta X_s$ with $H_{r\!f}$ (see fig.~\ref{fig2}c).
We have managed to overcome in part the first drawback of the model by
introducing an effective local flux density $B_{eff}$, which interacts
with the $r\!f$ current. When both the junction and grains are in the
Meissner or mixed state, the magnetic flux is quasi-homogeneously
distributed throughout the region to which it penetrates, and $B_{eff}=B$.
However, in the mixed state of the WL only, the flux will be concentrated
inside the junctions due to screening currents in the grains, and hence,
in the WL $B_{eff}>B$. The ratio $B_{eff}/B$  should increase with $B$ and
reach a maximum at $B_{eff}=\mu_0 H_{c1}$ ($H_{c1}$ is the lower critical
field of the grains) after which it should decrease rapidly.  Adopting a
simple function for $B_{eff}/B(B)$ which possesses the properties
specified above (we took the Gaussian function) we have managed to get an
excellent fit to our $\Delta X_s(H_{r\!f})$ data (see fig.~\ref{fig2}a),
but we failed to reproduce the $\Delta R_s$ field dependence (see
fig.~\ref{fig1}a).  Moreover, such a model can not reproduce the decrease
in $\Delta X_s$ with $H_{r\!f}$ observed for TF3 sample at high
temperatures (see fig.~\ref{fig2}c).

Another model applicable to our results is the model of
Eliashberg~\cite{Eliash} for superconductivity stimulated by high
power microwave irradiation. As shown by Eliashberg~\cite{Eliash},
in a superconductor with a homogeneous order parameter distribution,
microwave radiation of a certain power can induce a new quasiparticle
distribution function with an increased gap, which in turns leads to an
enhancement in superconducting properties. A similar effect is predicted
by the Aslamazov-Larkin (AL) theory~\cite{Aslar} for inhomogeneous WL,
due to the radiation-induced diffusion of quasiparticles out
of the junction region which occurs for a certain level of microwave power.
Since the AL theory, contrary to the Eliashberg model, predicts a
suppression of the order parameter at low $r\!f$ fields, we can exclude
this mechanism immediately, since we observe a {\it reduction\/} of $R_s$
at the lowest fields (see fig.~\ref{fig1}b,c).
As regards to the Eliashberg theory, it predicts a decrease of the
stimulation effect with lowered temperature, and a suppression of
superconductivity by a static magnetic field~\cite{Dmitr}. In fact, for
sample TF2 we see that the decrease in $R_s$ with $H_{r\!f}$ is reduced
with increasing temperature and completely disappears at high $T$, whereas
for sample TF3 the decrease in $R_s$ is observed only at high $T$
(see fig.~\ref{fig1}c).  Moreover, additional experiments
performed by us in low $dc$ magnetic fields $H_{dc}\leq 12$~mT (to be
published elsewhere), showed that while for sample TF2 $H_{dc}$
causes an even more pronounced decrease in $R_s$, for sample TF3 the $dc$
field {\it always leads to an enhanced\/} $R_s$~\cite{orig}. However, in
accordance with~Ref.[\onlinecite{Eliash}], stimulation of
superconductivity is expected only in highly uniform narrow and thin
superconducting channels with a homogeneous order parameter and microwave
field distribution, which is hardly the case for our wide and
``quasi-bulk" samples.

Finally, the third mechanism which may account for our results is the
recovery of superconductivity  due to the field-induced spin alignment of
magnetic impurities which are likely to be present in most HTS
(particularly in $Y\!BaCuO$)~\cite{Kres1}.
Magnetic impurities are a source of Cooper pair breaking due to the
spin-flip scattering process.  However, at low temperatures the decrease
in thermal motion leads to the appearance of spin-spin correlation of the
impurity atoms which becomes strong and may frustrate the spin-flip
scattering. An external magnetic field also leads to ordering via
alignment of the magnetic impurity spins and hence, can also lead to a
reduction of pair breaking.

\begin{figure}[t]
\def\epsfsize#1{0.4}
\vspace*{-3.0 true cm}
\centerline{\epsfbox{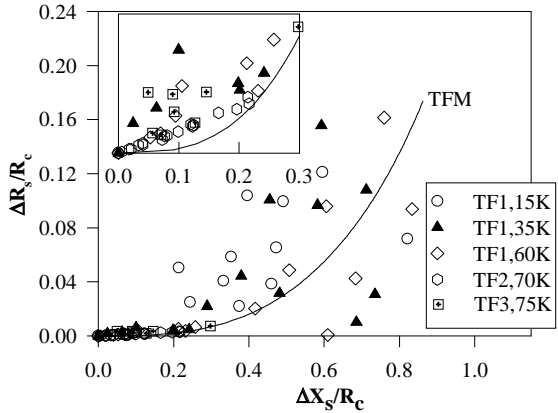}}
\vspace*{-4.5 true cm}
\caption{Parametric plot of $\Delta R_s/R_c$ vs.\ $\Delta X_s/R_c$ for
various samples at different temperatures (specified in the
figure). The solid line is a fit to the two-fluid model discussed in the
text. The insert shows the low power data on an expanded scale.}
\label{fig4}
\end{figure}

Recently Hein et al.~\cite{Hein} observed a correlated reduction of $R_s$
and $\lambda$ in both $dc$ and $r\!f$ fields of the same order of
magnitude ($\leq 20$~mT).  They performed an analysis of the function
$\Delta R_s/R_c(\Delta X_s/R_c)$ (where
$R_c=\sqrt{\omega\mu_0/2\sigma_n}$, and $\sigma_n$ is the normal electron
conductivity) in terms of the two-fluid model (TFM) and found that their
data collapsed onto a single TFM curve.  In addition,
the conductivity ratio $y=\sigma_1/\sigma_2$ (where $\sigma_1$ and
$\sigma_2$ are quasiparticle and superfluid conductivities, respectively)
was found to decrease with increased magnetic field, which was attributed
to the field-induced reduction of pair breaking.  The major difference
between our results and those of Hein et al.~\cite{Hein} is that they did
not observe a reduction of $R_s$ with $H_{r\!f}$ without an accompanying
reduction of $\lambda$; but at the same time they observed a reduction of
$\lambda$ with $R_s$ being almost independent of $H_{r\!f}$. In contrast,
we observed a reduction of $R_s$ for a monotonously increasing $\lambda$,
and moreover, only in rare cases did we observe a decrease in $R_s$
correlated with a decrease in $\lambda$ (see fig.~\ref{fig1}c and
fig.~\ref{fig2}c). In addition, a similar analysis based on the TFM was
performed by us which showed rather poor scaling of our $\Delta
R_s/R_c(\Delta X_s/R_c)$ data, as plotted in fig.~\ref{fig4}. One further
distinctive feature of our data compared to those of Hein et
al.~\cite{Hein} is the significant discrepancy (up to several times) in
the $R_c$ values extracted from the $\Delta R_s$ and $\Delta X_s$ data
(see table~\ref{tbl:1}). Besides, our conductivity ratio
$y=\sigma_1/\sigma_2$, extracted from the fitting to the TFM curve,
was found to increase with $H_{r\!f}$, rather than decrease, as expected
for the mechanism of the impurity spins alignment~\cite{Kres1}.
Nevertheless, our preliminary measurements in weak $dc$ magnetic fields
($\leq 12$~mT) in field cooled regime at constant $H_{r\!f}$ showed a
decrease of $R_s$ and $\lambda$ with $H_{dc}$ for samples TF1 and TF2 (to
be published elsewhere). This suggests that magnetic impurities may play a
significant role in our samples and might also affect our nonlinear
measurements.

\begin{table}[t]
\caption[] {Two fluid model fitting parameters of $\Delta R_s/R_c$ vs.\
$\Delta X_s/R_c$ dependences for various samples at different
temperatures. Here $y(H_{r\!f}=0)$ and $y(H_{r\!f}^{on})$ are the
conductivity ratios ($y=\sigma_1/\sigma_2$) at low and high microwave
powers, and $R_c(\Delta R_s)$ and $R_c(\Delta X_s)$ are the $R_c$ values
extracted from $\Delta R_s(H_{r\!f})$ and $\Delta X_s(H_{r\!f})$ data,
respectively.}
\begin{tabular}{p{2.2 cm}cccc}
\multicolumn{1}{c}{\hspace{-0.6 true cm} Sample}
& \hspace{-0.9 true cm} $y(H_{r\!f}=0)$ & \hspace{-0.1 true cm}
$y(H_{r\!f}^{on})$  & \hspace{-0.2 true cm} $R_c(\Delta R_s)$, m$\Omega$ &
$R_c(\Delta X_s)$, m$\Omega$ \\ \hline

TF1, 15 K & \hspace{-0.6 true cm} 0.183 & 0.173 & 0.103 & 0.046  \\ \hline
TF1, 35 K & \hspace{-0.6 true cm} 0.028 & 0.123 & 0.215 & 0.084 \\ \hline
TF1, 60 K & \hspace{-0.6 true cm} 0.050 & 0.111 & 0.700 & 0.352 \\ \hline
TF2, 70 K & \hspace{-0.6 true cm} 0.008 & 0.012 & 0.021 & 0.004 \\ \hline
TF3, 75 K & \hspace{-0.6 true cm} 0.010 & 0.022 & 0.0008 & 0.0003\\
\end{tabular}
\label{tbl:1}
\end{table}
\bigskip
\vspace*{-0.4 true cm}

We proceed with an analysis of our high-temperature data for increasing
$\Delta R_s(H_{r\!f})$ and $\Delta X_s(H_{r\!f})$ in terms of the
$r$-parameter~\cite{Golos2,Halbr5}. It is essential to consider
not only the $r$-value, but also the power dependence of the
impedance~\cite{Velich1}. A mechanism, such as the response of Josephson
vortices, for which $\Delta R_s$,$\Delta X_s\sim H_{r\!f}^n$ ($0.5<n<2$)
and the $r$-value is about unity, can be excluded immediately, since for
our experiments $r$ never exceeds 0.15. Moreover, the fit of our $r\!f$
field dependences with a function $\sim H_{r\!f}^n$ (see
fig.~\ref{fig1}b,c and fig.~\ref{fig2}b,c) has revealed uncorrelated
values of $n$ for  $\Delta R_s$ and $\Delta X_s$ data (while from
the theory~\cite{Halbr5} they should be the same), saying nothing
about an apparent departure of the $H_{r\!f}^n$ fit from $\Delta
X_s(H_{r\!f})$ data at high fields (fig.~\ref{fig2}b,c). The same
conclusion is valid for the heating of weak links ($r_{HE}<1$, $\Delta
R_s,\Delta X_s\sim H^2$) and the RSJ model~\cite{Halbr5} ($r_{RSJ}<1$,
$\Delta R_s$ increasing and $\Delta X_s$ oscillating with $H_{r\!f}$). A
value of the $r$-parameter consistent with our data could follow from
either uniform heating or intrinsic Ginzburg-Landau
nonlinearity~\cite{Golos2} (for both mechanisms $r<10^{-2}$), but the
$r\!f$ dependence $\sim H_{r\!f}^2$ is generally not observed for our
samples (except sample TF1, for which $\Delta R_s\sim H_{r\!f}^2$, but
$\Delta X_s$ is not $\sim H_{r\!f}^2$, see
fig.~\ref{fig2}a). Moreover, the $r$-value should increase with $T$ for
the above two mechanisms, while we see almost $T$-independent behavior at
high temperatures (fig.~\ref{fig3}b,c). Thus, our
high-temperature data which shows an increase in $\Delta R_s$ and
$\Delta X_s$ with $H_{r\!f}$ are apparently not explained by any of
the known theoretical models.

   In conclusion, in our experiments we appear to observe a
complicated interplay of several nonlinear mechanisms. At low
temperatures, the observed reduction in $R_s$ may arise due to the effect
of the magnetic impurity spins alignment by the $r\!f$-field,
while at higher $T$ stimulation of superconductivity by microwave
irradiation and vortex mechanisms may also come into play. However,
universal temperature- and sample-independent $H_{r\!f}$-dependence of the
penetration depth $\lambda$ (or, equivalently, the surface reactance
$X_s$) and similar values of $r\sim $0.01--0.06 for all the samples over a
broad temperature range ($35<T<75$~K) do not rule out the possibility that
all the observed features may arise due to one and the same mechanism, the
origin of which is not known at the moment. At the same time, absence of
correlation between $\Delta R_s(H_{r\!f})$ and $\Delta X_s(H_{r\!f})$,
for some of the samples particularly TF1, implies that
microstructure of the samples may interfere with the intrinsic behavior in
the nonlinear response.


\begin{references}
\bibitem{orig}
This suggests that the mechanisms of $R_s$ reduction
may be different for these samples.
\end{references}
\end{document}